\documentclass{aip-cp}

\usepackage[numbers,elide, sort&compress]{natbib}
\usepackage{rotating}
\usepackage{graphicx}
\usepackage{textcomp}
\usepackage{booktabs}

\usepackage{verbatim}
\usepackage{multirow,booktabs,colortbl,tabularx}
\usepackage[flushleft]{threeparttable}

\newcommand{\um}{$\mu$m }
\newcommand{\umnos}{$\mu$m}

\begin{document}

\title{Germanium Collimating micro-Channel Arrays For High Resolution, High Energy Confocal X-ray Fluorescence Microscopy}

\author[aff1]{David N Agyeman-Budu\corref{cor1}}
\author[aff2]{Sanjukta Choudhury}
\author[aff3]{Ian Coulthard}
\author[aff4]{Robert Gordon}
\author[aff3]{Emil Hallin}
\author[aff5]{Arthur R Woll}

\affil[aff1]{Department of Materials Science and Engineering, Cornell University, Ithaca, NY 14853 USA}
\affil[aff2]{Geol. Sciences, Univ. of Saskatchewan, 114 Science Place, Saskatoon SK S7N 5E2 Canada}
\affil[aff3]{Canadian Light Source, 44 Innovation Blvd, Saskatoon, SK S7N 5E2 Canada}
\affil[aff4]{Department of Physics, Simon Fraser University, Burnaby, BC, CA}
\affil[aff5]{Cornell High Energy Synchrotron Source, Ithaca, 14853 NY USA}

\corresp[cor1]{Corresponding author: da76@cornell.edu}

\maketitle

\begin{abstract}
Confocal x-ray fluorescence microscopy (CXRF) allows direct detection of x-ray fluorescence from a micron-scale 3D volume of an extended, unthinned sample. We have previously demonstrated the use of a novel collection optic, fabricated from silicon, that improves the spatial resolution of this approach by an order of magnitude over CXRF using polycapillaries. The optic, called a collimating channel array (CCA), consists of micron-scale, lithographically-fabricated arrays of collimating channels, all directed towards a single source position. Due to the limited absorbing power of silicon, the useful energy range of these optics was limited to fluorescence emission below about 10 keV. Here, we report fabrication of CCAs from germanium substrates, and demonstrate their practical use for CXRF up to 20 keV. Specifically we demonstrate a nearly energy-independent critical spatial resolution $d_R$ of 2.1$\pm$0.17 \um from 2-20 keV, as well as excellent background reduction compared to silicon-based CCAs throughout this energy range. Design details of the optic and background-reduction holder are described. Two versions of the optic are now available upon request at the beamline 20ID-B, Advanced Photon Source (APS) - Argonne National Laboratory.
\end{abstract}

\section{INTRODUCTION} \label{sect:intro}
Since its first practical demonstration just over ten years ago ~\cite{proost2002feasibility,Kanngiesser:2003nimb,janssens2004confocal,woll:2006appa,Woll:2008sic}, confocal x-ray fluorescence microscopy (CXRF) has become increasingly well known and commonly applied as a means of directly measuring 3D elemental composition in heterogeneous samples. CXRF is performed by scanning a 3D probe volume, defined by the overlapping foci of an excitation optic and a collection optic, through a sample of interest. As described in several recent reviews \cite{kanngieser_deep_2012,Perez:2010,Fittschen:2011}, the technique has been implemented at least at eight synchrotrons worldwide, as well as with lab-based sources \cite{Tsuji:2007,Perez:2011}, and for a wide variety of applications. 

Virtually all demonstrations of CXRF to date have employed a polycapillary optic as the collector. A key advantage of these optics for this application is their large angle of collection, upwards of 25 degrees, corresponding to 1\% of 4$\pi$ sterradians. However, a disadvantage of polycapillary optics is their spatial resolution, which has been limited to approximately 10 \um~at 10 keV. And since polycapillaries function on the principle of critical reflection, they exhibit severe chromaticity in both spatial resolution and sensitivity. The magnitude of this spatial resolution and its dependence on fluorescence energy are documented in many papers  \cite{wolff_performance_2009,kanngieser_deep_2012}, and have been explicitly cited as key limitations of the technique \cite{Mantouvalou20129,Nakazawa:2013,pemmer_bone:2013}.

\begin{figure}[ht]
   \includegraphics[width=.9\textwidth]{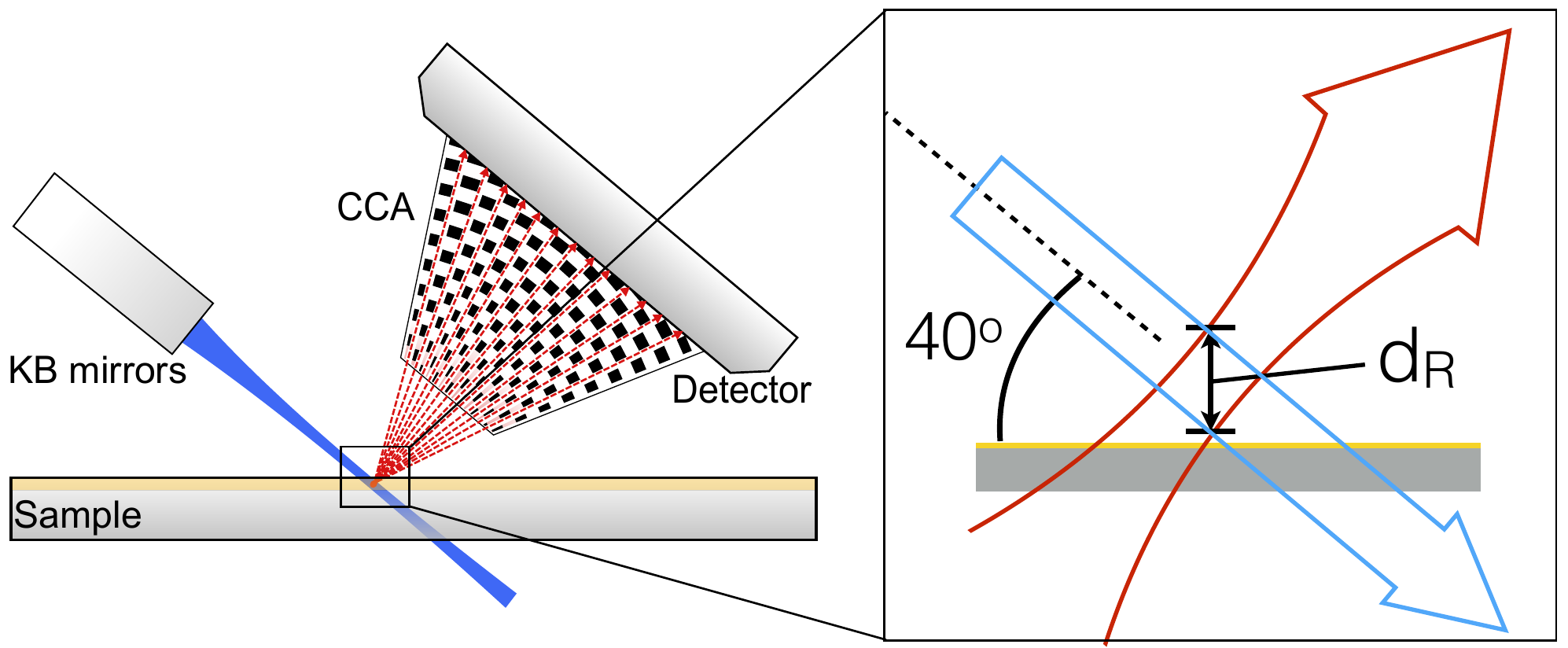}
 \caption{Schematic of the layout of the CXRF measurement geometry with a CCA optic, illustrating the formation of a probe volume by the overlapping foci of the KB mirrors and CCA, and the transmission of fluorescence x-rays from the probe volume, through the channels, and then to the detector. As described in Ref.\cite{woll2014} and below in the design section, the channels are formed by a staggered arrangement of pillars, giving rise to the distorted checkerboard pattern.}
 \label{fig:figure1}
%
%
\end{figure}

Recently, we demonstrated the first use of lithographically-etched channel arrays in silicon as a collection optic for CXRF\cite{woll_3d_2012,woll2014}, as well as their application towards 3D-resolved micro-XANES in ancient stained glass\cite{choudhury:2015}. The experimental configuration is shown schematically in Figure~\ref{fig:figure1}, illustrating that the channels function as an array of collimators. The fields of view of each collimating channel overlap in a single focal region, which is aligned to coincide with the focal position of the incident beam. The intersection of these two foci define a three-dimensional probe volume. Using this approach, we achieved an order-of-magnitude improvement in resolution when compared with previous demonstrations of CXRF using polycapillaries as the collection optic. As an example, Ref.\cite{woll2014} illustrates the a depth resolution $d_R$ (defined as the projected size of the probe volume normal to the surface of a sample as in Figure~\ref{fig:figure1}) of under 2 \um for 3-10 keV.  

The upper fluorescence energy limit of 10 keV in our prior demonstrations of CCAs for CXRF arises from the limited ability of silicon-based optics to suppress background x-rays from outside the probe volume. This limitation in energy corresponds to an inability to probe elements from a large fraction of the periodic table. The choice of silicon, despite this limitation, was driven by its compatibility with deep reactive ion etching (DRIE) which, in turn, is employed to maximize channel depth. Fortunately, germanium is also compatible with isotropic fluorine-based etching \cite{oehrlein:1991,darnon:2015}, and has substantially greater x-ray attenuation than silicon. Here, we demonstrate the use of germanium-based CCAs for CXRF, allowing energy-independent energy resolution from 2-20 keV. We also demonstrate that these optics should work well up to 30 keV, at which energy the attenuation of germanium is comparable to that of silicon at 10 keV, where silicon begins to fail. 

The plan of the paper is as follows: after reviewing the design constraints of collimating channel arrays, we describe critical details regarding their fabrication using DRIE. Next we describe a custom-designed holder to mount CCA optics to a Vortex EX detector and minimize background. Finally, we describe our experimental realization of CXRF at 20ID-B, as well as our key experimental result comparing the performance of Si- and Ge-based CCAs. 

\section{CCA DESIGN AND FABRICATION}\label{sect:design}

\subsection{Channel Design} \label{sect:channels}

As described in Ref.~\cite{woll_3d_2012}, collimating channel arrays are characterized by their width $w$, length $l$, depth $d$, working distance $f$, and the total number of channels $N$. An additional parameter, $\delta$, corresponds to the angular acceptance of the channels\footnote{Note that this definition of $\delta$ is a factor of 2 smaller than that defined in Ref.~\cite{woll_3d_2012}, and results in a corresponding change to Eq.~\ref{eq:res}}. In the limit that the channels are perfect collimators, $\delta=w/l$. With these definitions, we may write approximate values for the two principal figures of merit for a channel array, namely their spatial resolution:
\begin{equation} \label{eq:res}
r = w +\delta f,
\end{equation}
and their fractional solid angle of collection:
\begin{equation} \label{eq:efficiency}
\eta = \frac{N \delta d}{4 \pi (f+l)}.
\end{equation}
These equations illustrate that maximizing channel depth $d$ increases $\eta$ in direct proportion. They also show that any choice in channel length represents a compromise between minimizing $r$ and maximizing $\eta$: longer channels will generally reduce $\delta$ and hence $r$, but will also reduce $\eta$. An even more critical constraint on channel length is to achieve good background rejection: that is, $l$ must be long enough that x-rays originating outside the probe volume are absorbed. Absorption of x-rays is governed by the well-known Beer-Lambert-Bouguer Law, which for monochromatic radiation may be expressed as $I=I_0 \exp({-\tau/\alpha(E)})$, where $\tau$ is the thickness of material traversed by the beam, and $\alpha(E)$ is the energy-dependent attenuation length of that material. Ray-tracing simulations (not shown) suggest that the minimum channel length $l_{\textnormal{min}}$ for sufficient background reduction should be at least 15 times the attenuation length:
\begin{equation}  \label{eq:lmin}
l_{\textnormal{min}} \approx 15 \times \alpha(E).
\end{equation}
In addition to the constraints above, an upper limit on channel length is imposed by diffraction effects. Specifically, we expect diffraction-induced spreading of the photon field to result in absorption of photons that would otherwise be directed through a channel, reducing the efficiency $\eta$. From basic path-length considerations, we expect this to manifest above some maximum channel length $l_{\textnormal{max}}$:
\begin{equation} \label{eq:lmax}
l_{\textnormal{max}} \propto E w^2/2. 
\end{equation}

\begin{figure}[ht]
\includegraphics[width=0.8\textwidth]{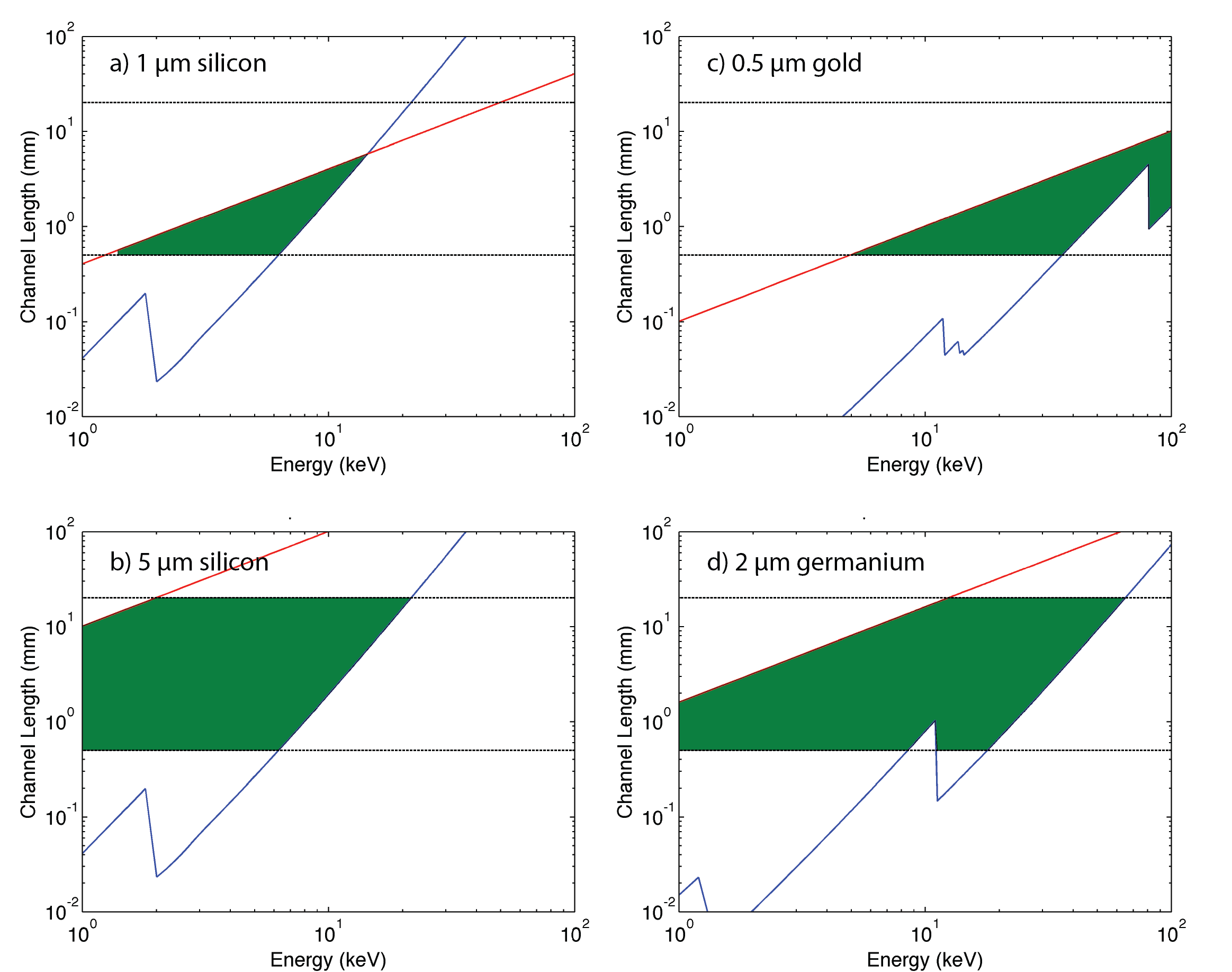}
\caption{Channel length vs. energy plots for four choices of channel width and absorber material: (a) 1-\um silicon channels, (b) 5-\um silicon channels, (c) 0.5-\um gold and (d) 2-\um germanium channels. In each plot, the lower line at high energy (blue online) corresponds to $l_{\textnormal{min}}=15\times \alpha(E)$ required for sufficient absorption of unwanted emission from the sample, and the upper line at low energy (red online) to $l_{\textnormal{max}}=E w^2/2$, above which diffraction-induced losses are expected to play a role. The overlap region (shown in green) shows the region in which channels are expected to perform well.} \label{fig:sca_design}
\end{figure}

Figure~\ref{fig:sca_design} provides visual representations of Equations \ref{eq:lmin} and \ref{eq:lmax} for several choices of channel width and absorbing material. In each plot, the red and blue lines correspond to $l_{\textnormal{min}}$ and $l_{\textnormal{max}}$, respectively. Horizontal dashed lines represent the practical, though somewhat arbitrary constraint that the channel length be between 0.5 and 20 mm. The green region in the plots represent the parameter space in which the channels are expected to have both sufficient background reduction and energy-independent efficiency $\eta$. Figures~\ref{fig:sca_design}a-b correspond closely to the optics described in both Refs.\cite{woll_3d_2012} and \cite{woll2014}, and demonstrate visually the behavior of these optics described above: at energies above about 10 keV, their absorbing power becomes insufficient to provide sufficient background reduction.

Figures \ref{fig:sca_design}c-d illustrate the benefits and limitations of fabricating channel arrays from materials other than silicon. For example, Figure~\ref{fig:sca_design}c suggests that, although fabrication of CCAs from gold -- for example using electroforming techniques -- might allow 0.5-\um wide channels to have sufficient absorbing power at a length of only 0.5~mm, diffraction losses may still prevent such a device from operating efficiently at energies \textit{below} about 10 keV. Alternatively, Figure~\ref{fig:sca_design}d shows that 2-\um$\times$ 4-mm channels fabricated from germanium should work well over a large energy range, from 3 to 30 keV. 

\subsection{Mask Design} \label{sect:mask4}

The original CCA design described in Ref.\cite{woll_3d_2012} suffered from a very small working distance $f$ of 0.2 mm and limited channel depth $d$ of 30-50 \umnos. Ref.\cite{woll_3d_2012} also shows that, owing to strong diffuse reflection by the (continuous) channel walls, the angular selectivity $\delta$ was much larger than the geometric collimation $w/l$. Ref.~\cite{woll2014} demonstrates that forming channels from a staggered array of pillars, as illustrated in Figure~\ref{fig:figure1}, decreases $\delta$. This allowed CCAs with 1-\um channel-width and 0.5 mm  working distance to retain the same spatial resolution as demonstrated in Ref.\cite{woll_3d_2012}. An additional feature of the channels described in Ref.\cite{woll2014} was that the pillars forming the channels were tapered away from the beam direction in order to further reduce reflections. Further analysis suggested that such tapering may not be necessary and, moreover, that \textit{refraction} from the tapered pillars can increase $\delta$, thus \textit{harming} spatial resolution $r$. 

\begin{figure}[ht]
\includegraphics[width=0.9\textwidth]{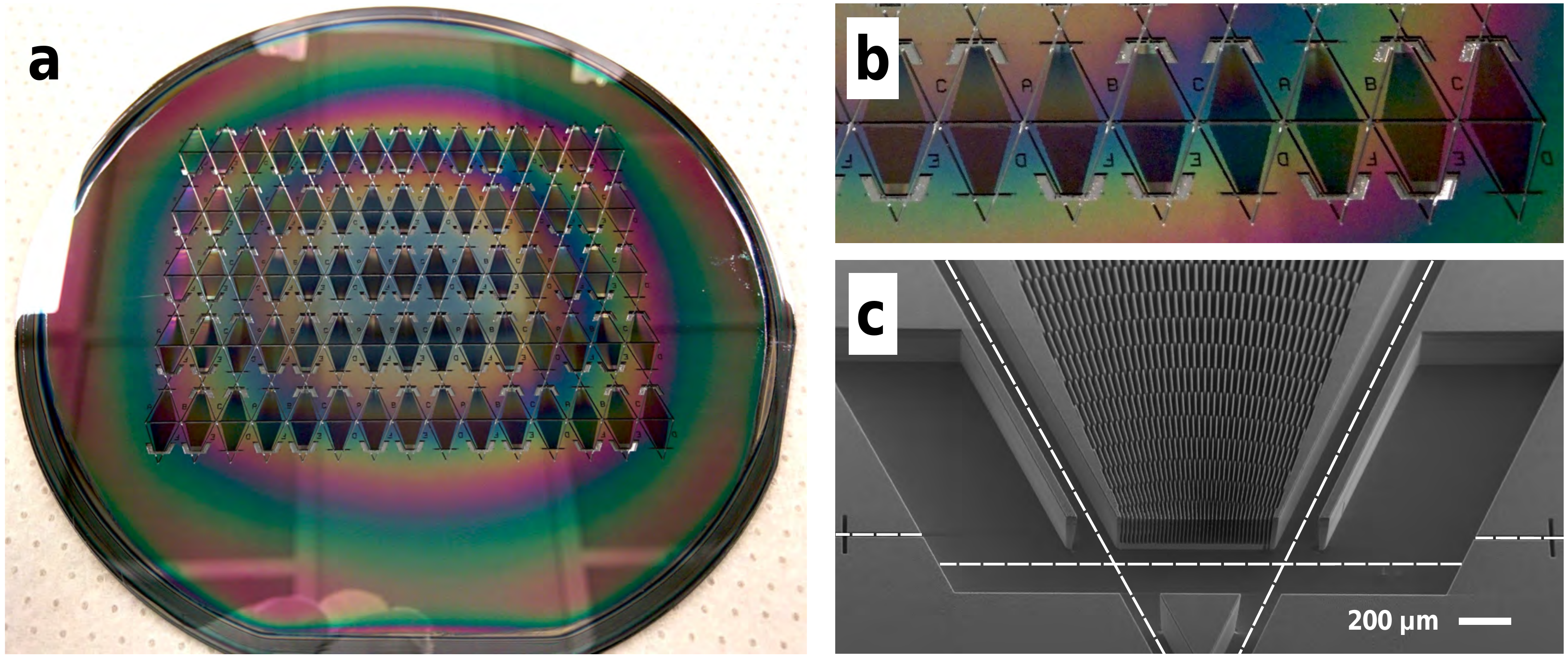}
\caption{(a) Photograph of a silicon substrate etched using the mask described in the text. The mask pattern comprises six distinct optics, arranged in a 3$\times$2 array, is repeated on a 5$\times$5 grid. (b) Zoomed in view of a portion of (a). (c) Scanning electron micrograph of optic F in (a), indicating the lines along which the wafer is diced to release optics from the wafer.} \label{fig:wafer_and_sem}
\end{figure}

Based on these observations, we designed a new mask, and used it to fabricate new optics from both silicon and germanium substrates. Figures~\ref{fig:wafer_and_sem}a-b show photographs of a silicon wafer etched with this mask. Although the mask comprises six different optic designs (labeled A-F, see Figure~\ref{fig:wafer_and_sem}b), this paper focuses entirely on optic A. This optic has a working distance of 1.0 mm, and comprises 121, 2-\um channels separated by 5 milliradians. The channels vary in length from 4 mm to 4.2 mm, increasing with channel angle from the optic axis. Although it is not further discussed in this paper, we note that optic D is similar to optic A, but comprising 75, 5-\um channels separated by 8 milliradians and a working distances of 1.5 mm. Optics A and D are available to general users in both silicon and germanium at beamline 20ID-B, APS. In contrast to the optics described in Ref.\cite{woll2014}, the walls of each pillar in optics A and D are not tapered inward, but are parallel to each channel.

Each channel array in Figures~\ref{fig:wafer_and_sem}a-b is inscribed within identically-shaped triangular regions, each with an apex angle of 45\textdegree and a base-to-tip height of 5 mm. This uniformity allows the different CCAs to be interchangeably mounted to a single holder, described further below. When the wafer is diced to separate CCAs from the wafer, each optic takes on the shape of a trapezoid. Figure \ref{fig:wafer_and_sem}c shows an SEM image of one of optics defined by the mask, also indicating the lines along which the wafer is diced. A new feature of this mask design in contrast to prior versions is the inclusion of solid side walls adjacent to the pillars. These walls are visible in Figure~\ref{fig:wafer_and_sem}c, and serve both to mechanically protect the walls and to aid in background reduction. 

\subsection{Fabrication} \label{sect:fab}

CCA optics are fabricated using a combination of standard photolithography and DRIE, the details of which are described in Ref.\cite{woll_3d_2012}. However, the staggered pillar arrangement described above and in Ref~.\cite{woll2014}, which enabled channels to be etched to much greater depth, also necessitated the introduction of a hard mask rather than photoresist. For this purpose we evaporated a 50-nm Al$_2$O$_3$ film onto 2 \um SiO$_2$, itself deposited using a GSI Plasma Enhanced Chemical Vapor Deposition (PECVD) system. This mask was employed for both silicon and germanium-based optics.

\begin{figure}[ht]
\includegraphics[width=0.9\textwidth]{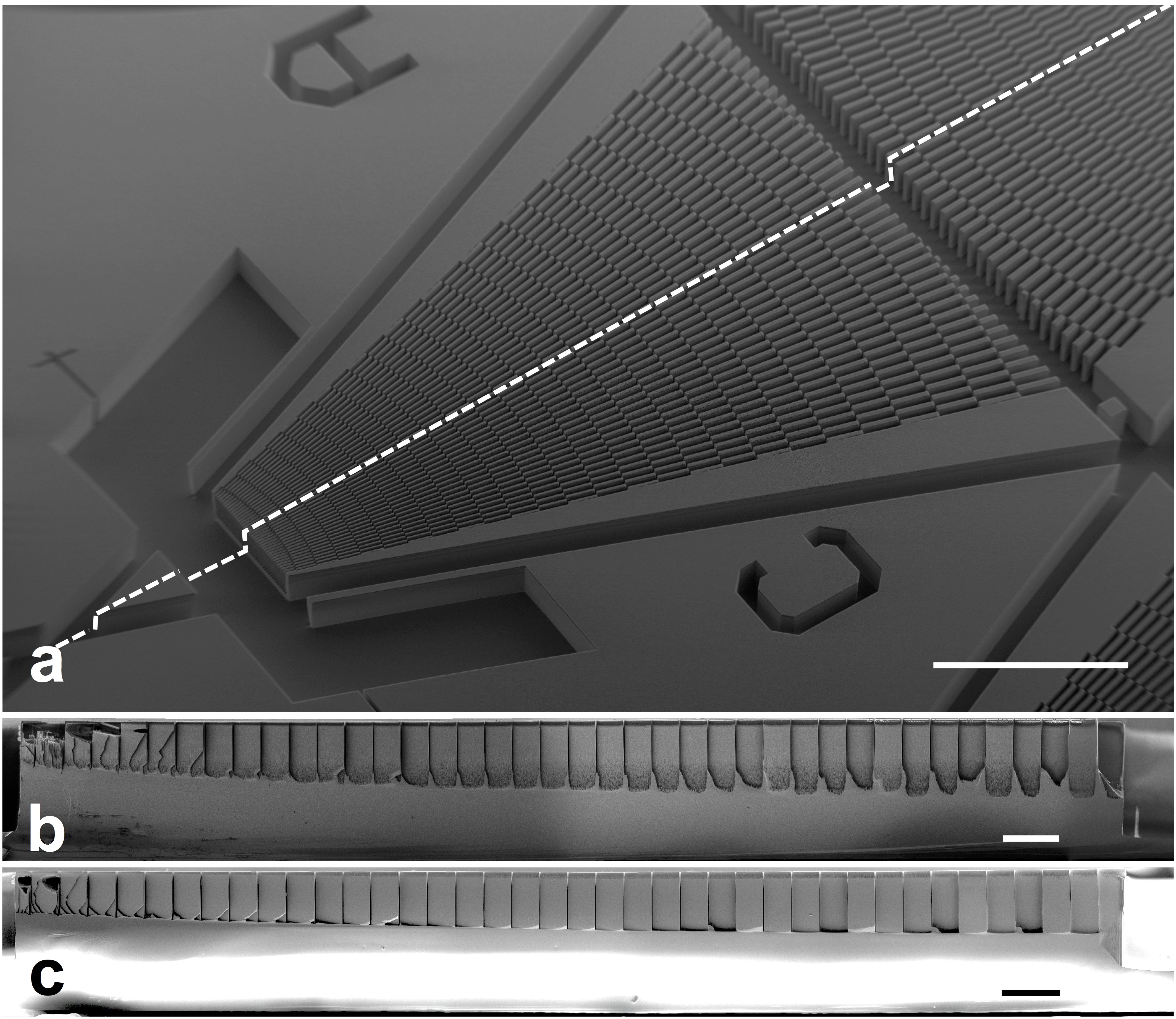}
\caption{(a) Aerial view of the CCA with the dashed line indicating the position of the cross-sectional views along the optic axis shown below in (b) and (c). The scale bar is 1 mm (b) CCA optic etched with a standard procedure illustrating one manifestation of aspect-ratio-dependent etching. (c) Cross-section view of a CCA optic etched with varying etch parameters that vary with depth in order to preserve channel dimensions. Scale bars for (b) and (c) are 200 $\mu$m  } \label{fig:morphing}

\end{figure}

The staggered pillar arrangement increases the effective trench width for etching without impacting the channel size. In this way, staggered pillars mitigate the effect of aspect ratio dependent etching (ARDE). For example in optic A described above, the pillars defining the 2 \um channels at the entrance to the optic are 7 \um apart. However, ARDE still introduces challenges to making ideal CCAs. First, it can result in a change in feature shape as a function of depth. Second, staggered pillars introduce the complication that the trench width changes along the channel. Figure~\ref{fig:morphing}a shows an SEM of the whole of optic C, and clearly illustrates the changing lateral feature size and trench width from the entrance towards the exit of the optic. We expect this change in trench width to result in a changing channel depth. 

To evaluate these effects, we have a series of optics by cleaving them along the dashed line indicated in Figure~\ref{fig:morphing}a in order to view them as in Figs.~\ref{fig:morphing}b-c. We note that the optics in Figs.~\ref{fig:morphing}b-c were produced with a different mask than that used for the optic in Figure~\ref{fig:morphing}a. They nevertheless illustrate the issues described above.  First, the etch depth in Figure~\ref{fig:morphing}b increases from 175 \um to 290 \um from the optic's entrance to its exit. This difference clearly arises from the change in trench width, which in this case varied from $\approx$ 3.6 \um to 25 \umnos. Second, the imperfect, rounded appearance of the bottom of the trenches in Figure~\ref{fig:morphing}b indicates a change in channel morphology with increasing depth -- possibly a decrease in channel width.

The first of these effects, an increasing channel depth as a function of distance from front to the back of the optic, does not necessarily harm the function of the optic, provided that the channel depth at the front of the optic does not limit the total solid angle of collection. However, loss of channel width does impact CCA performance. Hence we investigated modifying the DRIE etch recipe to mitigate against ARDE-induced effects.

DRIE consists of three steps: deposition of a teflon-like polymer thin film for sidewall passivation of the trench, a highly directional etch step to remove the polymer at the bottom of the trench, and an isotropic silicon etch. The PlasmaTherm Versaline Deep Silicon Etch tool used for this work refers to these steps as ``Dep", ``Etch A", and ``Etch B" respectively, and each complete sequence of these steps is a ``loop". A recipe specifically developed for etching deep features in silicon with this tool is described by table \ref{tab:etch_pars}, which lists the parameters for each step.

\begin{table}[ht]
\caption{Process parameters defining the ``standard recipe" for a single etch loop of a DRIE process used to create CCAs. When ``Start" and ``End" parameters differ within a single portion of the loop, that parameter varies linearly during that step.
\label{tab:etch_pars}}
\centering
  \begin{threeparttable}
	\begin{tabular}{lccccccccc}
		\toprule
			& \multicolumn{2}{c}{Dep} & & \multicolumn{2}{c}{Etch A}  & & \multicolumn{2}{c}{Etch B}  \\
			\cmidrule(lr){2-3} \cmidrule(lr){5-6} \cmidrule(lr){8-9}

			Parameter  & Start & End  &   & Start & End  &   & Start & End &  \\
			\midrule
            Ar flow rate (sccm\tnote{\textdagger} ) & 10   & 10   &  & 10   & 10   &  & 10   & 10   &  \\
            Bias RF Voltage set point (V)		    & 10   & 10   &  & 400  & 425  &  & 10   & 10   &  \\
            C$_4$F$_8$ flow rate (sccm) 		    & 150  & 150  &  & 150  & 150  &  & 150  & 150  &  \\
            ICP RF F. power set point (W)	  	    & 2000 & 2000 &  & 1500 & 1500 &  & 2000 & 2000 &  \\
            Pressure (mTorr) 			   		    & 25   & 25   &  & 20   & 20   &  & 40   & 40   &  \\
            Process set point (sec) 	 	 	    & 2.3  & 2.0  &  & 2.0  & 2.0  &  & 2.0  & 2.0  & \\
            SF$_6$ flow rate (sccm)		  		    & 39   & 39   &  & 32   & 32   &  & 32   & 32   & \\
            Throttle valve position set point (\%)  & 100  & 100  &  & 100  & 100  &  & 150  & 150  &  \\
\bottomrule
\end{tabular}
 \begin{tablenotes}
             \item[\textdagger] Standard cubic centimeters per minute
        \end{tablenotes}
  \end{threeparttable}
\end{table}

The optic in Figure~\ref{fig:morphing}b was etched with this recipe for 1050 loops. The optic represented in Figure~\ref{fig:morphing}c was etched for the same total number of loops as that in Figure~\ref{fig:morphing}b, but using a modified approach in which  the parameters of the standard recipe are modified continuously over the course of the etch. This modification is referred to as ``morphing''. In this case, the 1050 etch loops were divided into three parts consisting of 400, 400, and 250 loops each. The first part used parameters identical to those in table \ref{tab:etch_pars}. In parts two and three, the Bias RF voltage set point and process time were modified as indicated in table~\ref{tab:moph_pars}. The improvement of this modification on the etch quality is clearly illustrated by Figure~\ref{fig:morphing}c, in which the bottom-etch of the trenches are square.

In future, we anticipate fabricating CCAs using the morphing approach. However, we have not yet optimized it for use on germanium. The optics used for the x-ray characterization described below were thus fabricated using the standard recipe represented by table~\ref{tab:etch_pars}. The silicon optic was etched for 1121 loops, while the germanium optic was etched for 720 loops.

\begin{table}[ht]
\caption{Changes in etch parameters for the 2nd and 3rd parts of the morphing recipe. \label{tab:moph_pars}}
\centering
  \begin{threeparttable}
      \begin{tabular}{ccccccccccc}
          \toprule
          & & \multicolumn{2}{c}{Dep} & & \multicolumn{2}{c}{Etch A}  & & \multicolumn{2}{c}{Etch B}  \\
          \cmidrule(lr){3-4} \cmidrule(lr){6-7} \cmidrule(lr){9-10}

          Part\tnote{\textdagger} & Parameter & Start & End  &   & Start & End  &   & Start & End &  \\
          \midrule

          \multirow{2}{*}{2} & \multirow{1}{*}{Bias RF Voltage set point (V)} & (10) & (10) &  & 425   & 450   & & (10)  & (10)  \\ 
          & \multirow{1}{*}{Process set point (sec)}						  & 2.0  & 1.8 	&  & (2.0) & (2.0) & & (2.0) & (2.0)  \\ \\
          \multirow{2}{*}{3} & \multirow{1}{*}{Bias RF Voltage set point (V)} & (10) & (10) &  & 450   & 450   & & (10)  & (10) \\ 
          & \multirow{1}{*}{Process set point (sec)}						  & 1.8  & 1.8 	&  & (2.0) & (2.0) & & 2.0   & 1.8  \\ 
          \bottomrule
      \end{tabular}
      	 \begin{tablenotes}
             \item[\textdagger] The 1st part is Table \ref{tab:etch_pars}.
        \end{tablenotes}
  \end{threeparttable}
\end{table}

\section{X-RAY CHARACTERIZATION}\label{sect:exp}

\subsection{Experimental Details}

%
The resolution and high energy performance of CCA optics described above were tested at the Advanced Photon Source, beamline 20ID-B. For the measurements described below, an incident beam energy of 20.1 keV was selected using a Si double crystal (111) monochromator and focused by a set of Kirkpatrick-Baez (KB) mirrors. The unattenuated flux in this spot was 4.3$\times 10^{10}$ photons/second, measured using a helium-filled ionization chamber downstream of the sample position. The beam size was measured, at 12 keV, by scanning a 100 \um tungsten wire through the beam, and found to be 1.7 \um tall $\times$  2.2 \um wide. Depth resolution measurements at 20.1 keV, described below, suggest that the horizontal beam size is most likely smaller than this value. The precise vertical beam size has no bearing the analysis described here. A 3D sample translation stage provided coarse and fine motion, enabling sub-\um control, and was arranged such that the sample surface was 40\textdegree~from the incident beam as shown in Figure~\ref{fig:figure1}. The confocal geometry was arranged by mounting CCA optics directly onto a single element silicon-drift Vortex EX detector (Hitachi High-Technologies Science America, Inc., Northridge, CA) positioned at 90\textdegree~to the incident beam. The detector was also placed on a 3D motorized stage and included sub-\um control only along the detector axis.  

\begin{figure}[ht]
\includegraphics[width=0.9\textwidth]{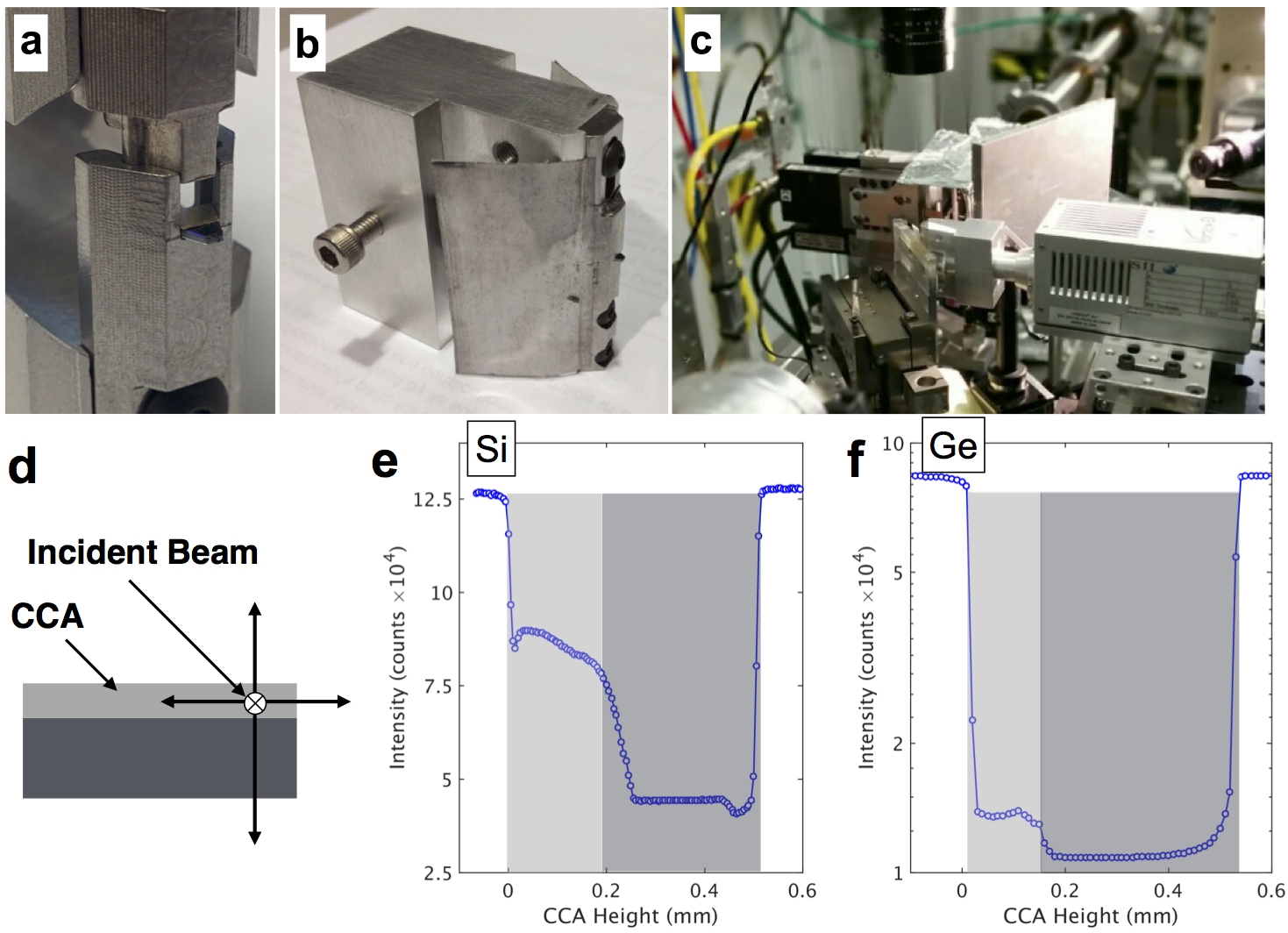}
\caption{(a) Photograph of the custom optic holder showing a close up view of the interlocking steel jaws with the CCA optic and silver foil mounted in place (b) Assembled optic holder showing lead tape on the sides to further reduce background. (c) Photograph of a CXRF setup at ID20, showing the holder in (b) mounted onto the Vortex EX detector. (d) schematic illustration of vertical and horizontal scans to align optic with respect to the incident beam. Light and dark gray regions correspond to etched and unetched portion of the substrate. (e) Vertical alignment scan of a Si-based ccA, showing transmitted intensity as a function of CCA height. (f) Vertical alignment scan as in (e) but of a Ge-based CCA.} \label{fig:holder_and_id20}
\end{figure}

From our prior experience aligning CCAs for confocal XRF, the most serious practical difficulty is elimination of background. Generally, background x-rays reach the detector from source points outside the probe volume, and traverse paths just around the optic or through other gaps in shielding. To mitigate this challenge, a custom CCA holder was designed and fabricated, and is shown in various stages of use in Figures~\ref{fig:holder_and_id20}a-c. The holder comprises three parts: an aluminum adaptor that mates to the snout of the Vortex detector, and two small pieces that mount to the adaptor and form interlocking jaws in which the CCA is clasped. The two jaws, machined by Protolabs, Inc (Maple Plain, MN), are designed to block most lines of sight between the sample and detector except those through the optic, and machined from 1018 steel in order to block x-rays effectively up to 20 keV. Figure~\ref{fig:holder_and_id20}a shows the front of the CCA holder, illustrating the assembly with a CCA in place but with the top jaw raised. The photograph shows one of two 100-\um silver foils (Alfa Aesar, Ward Hill, MA) placed immediately above and below the CCA, which provide malleable seals between the optic and jaws.  When the detector face is touching the back face of the jaws, it is 3.5 mm behind the CCA and 8.5 mm from the CCA focus. At this distance, the 35\textdegree~fan of x-rays traversing the optic spans approximately 5 mm across, and so can be fully captured by the Vortex SDD chip, which is 6 mm wide at mid-height\footnote{The Vortex SDD chip in this model detector is hexagonal in shape, with major and minor axes of the useful chip area of approximately 6 mm and 5.3 mm, respectively\cite{Feng_SPIE:2004}. In single-element vortex detectors, the major axis is aligned horizontally with respect to the detector housing\cite{Valeri_email:2014}.}. Since the holder does not completely block background, additional lead tape (see Figure~\ref{fig:holder_and_id20}b) is attached to the holder. Finally, Figure~\ref{fig:holder_and_id20}c shows a view of the experimental setup at ID20, including the Vortex detector and CCA holder as used.

Alignment of CCA optics as well as the resolution and background rejection tests described below were carried out with the aid of a metallic, multilayer film sputter deposited onto a silicon substrate. The film consists of approximately 10 nm of Cr, followed by approximately 100 nm each of Mo, Nb, Zr, and Au. CXRF alignment is greatly aided by placing a video camera directly above and looking down onto the setup (see Figure~\ref{fig:holder_and_id20}c), such that the distance between the optic and sample can be closely monitored. 

The alignment procedure is as follows: First, the CCA is coarsely aligned by scanning it through the incident beam both vertically and horizontally as indicated in Figure~\ref{fig:holder_and_id20}d, while monitoring the beam downstream of the optic using an ionization chamber. The vertical scan will exhibit enhanced transmission when the beam passes through the etched portion of the substrate, indicating the height and extent of the channels. Examples of such scans through the Si- and Ge-based CCAs are shown in Figs.~\ref{fig:holder_and_id20}e-f, which indicate approximate etch depths of 190 \um and 150 \um for these two optics, respectively. This information is used to approximately center the beam vertically with respect to the channels. 

Next, the front edge of the CCA is scanned transversely into the incident beam. The onset of reduced transmission in this scan serves to locate the edge, allowing it to be precisely translated away from the beam by the (well known) working distance of 1.0 mm for optic A. Finally, the thin metallic film sample is translated in $\approx$2 \um increments through the expected probe volume position, while monitoring the fluorescence intensity obtained at the detector. This translation corresponds to moving the source point along the incident beam. When it traverses the probe volume, it ideally produces a conspicuous, localized rise in XRF intensity at the detector. However, the fact that the optic is aligned only coarsely can reduce the peak intensity and degree of localization. If present, high background can also obscure this peak. Once some intensity from the probe volume is found, the precise horizontal distance between the incident beam and detector/CCA optic assembly is refined to maximize intensity and complete the alignment.

Although this paper focuses on the resolution $r$ and background reduction of germanium-based optic, we note that the efficiency $\eta$ of CCAs based on staggered channels was characterized in Ref.~\cite{woll2014}, and found to be commensurate with Eq.~\ref{eq:efficiency}, provided the angular selectivity $\delta$ is known. Measurements (not shown) of the angular selectivity of germanium-based channels of optic A indicate an angular selectivity of 1 milliradian, a factor of 2-3 smaller than obtained in Ref.~\cite{woll2014}. Using Eq.~\ref{eq:efficiency} with this value of $\delta$ leads to $\eta = 0.04\%$. This value is smaller than previously obtained, in part due to the improved (\textit{smaller}) angular selectivity. Nevertheless, this value of $\eta$ is sufficient to obtain detector-limiting XRF intensities from many samples, as illustrated below. 

\subsection{Results} \label{sect:results}

\begin{figure*}[ht]
\centering
  \includegraphics[width=\textwidth]{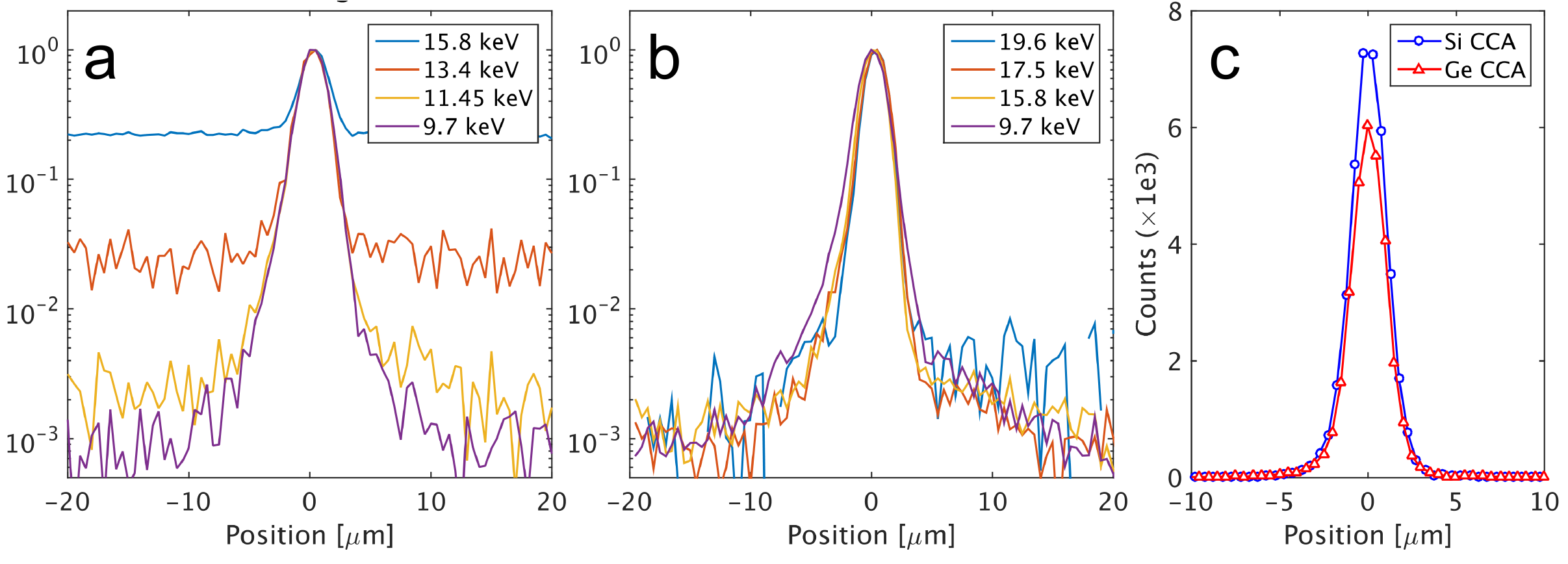}
  \caption{(a) Intensity vs. depth of Au L$\alpha$ (9.7 keV), L$\beta$ (11.45 keV), and L$\gamma$ (13.4 keV) and Zr K$\alpha$ (15.8 keV) emission lines obtained by scanning a multilayer metal film stack through a confocal XRF probe volume, using a silicon CCA described in the text. Intensities are normalized by their maxima to emphasize background.(b) As in (a) but using a germanium CCA as the collection optic and showing intensities of Au L$\alpha$(9.7 keV), Zr K$\alpha$ (15.8 keV), Mo K$\alpha$ (17.5 keV), and Mo K$\beta$ (19.6 keV) emission lines. (c) Intensity comparison of AuL$\alpha$ fluorescence with the Si and Ge CCAs in (a) and (b), normalized to the same incident intensity of 1.6$\times10^9$ photons/second incident intensity.  The maximum intensity difference is comparable to the etch depth difference from the results in Figures~\ref{fig:holder_and_id20}~e-f. \label{fgr:depth_scans}}
\end{figure*}

The multilayer metal film described above and used for alignment was also used to characterize both the resolution and background rejection properties of germanium-based CCA optics. In particular, it was aligned and scanned through the CXRF probe volume along its surface normal. Results of this scan using the silicon and germanium CCAs are shown in \ref{fgr:depth_scans}a-b, which show the XRF intensity of a subset of fluorescence lines originating from the film. In order to emphasize the background contamination -- intensity that is detected even when the film is translated outside the probe volume -- the data are shown on a log scale, and each profile has been normalized by its maximum. The difference is striking: although the Au L$\alpha$ plots in Figures \ref{fgr:depth_scans}a-b are nearly identical, the background level in Figure~\ref{fgr:depth_scans}a clearly increases with increasing energy, to $\approx$0.2\% of the maximum intensity for Au L$\beta$ at 11.45 keV, 6\% of the maximum at 13.4 keV, and is 25\% at 15.8 keV. By contrast, the identical scan with a germanium optic (Figure~\ref{fgr:depth_scans}b) exhibits nearly identical background levels for all energies. The lines plotted in Figure~\ref{fgr:depth_scans}b illustrate that this achromaticity persists up to 19.6 keV, corresponding to Mo K$\beta$ emission. The apparent,  increased background to the right side of the Mo K$\beta$ peak is an artifact of the lower intensity of this peak and, possibly, from Compton scattering background from the substrate. 

Figure \ref{fgr:depth_scans}c shows the same Au L$\alpha$ intensity distribution obtained from the scans shown in Figs.~\ref{fgr:depth_scans}a-b, but on an linear scale and normalized to an incident intensity of $\approx 1.6 \times 10^9$ photons per point, obtained by attenuating the beam with 40 \um Molybdenum. The maximum intensity for the Si CCA is approximately 20\% larger than that obtained with the Ge CCA. This difference is commensurate with the difference in channel height obtained from Figs.~\ref{fig:holder_and_id20}e-f. 


Figures \ref{fgr:depth_scans}a-b thus demonstrates that the substitution of germanium for silicon results in strong reduction of background fluorescence arising from outside the probe volume and impinging on the optic. Considering Figure~\ref{fgr:depth_scans}a again, we note that the energy, 10 keV, at which the silicon CCA begins to fail, corroborates equation \ref{eq:lmin}. At 10 keV, $\alpha(E)=134~ \mu m$\cite{henke_x-ray_1993}, corresponding almost exactly to $2\textnormal{mm}/15$. We may thus anticipate that the germanium-based version of these optics should work up to the energy at which $\alpha(E)$ approaches 130 \um, namely 30 keV. We note, however, that such operation will likely require additional strategies for reducing background traversing around the optic. For example, 30 keV operation will require a different material for the jaws of the optic holder described above, since Fe is not sufficiently opaque. In addition, the Pb tape used here (see Figure~\ref{fig:holder_and_id20}b) will be insufficient to absorb x-rays penetrating seams in the holder.

\begin{figure}[ht]
\centering
  \includegraphics[width=\textwidth]{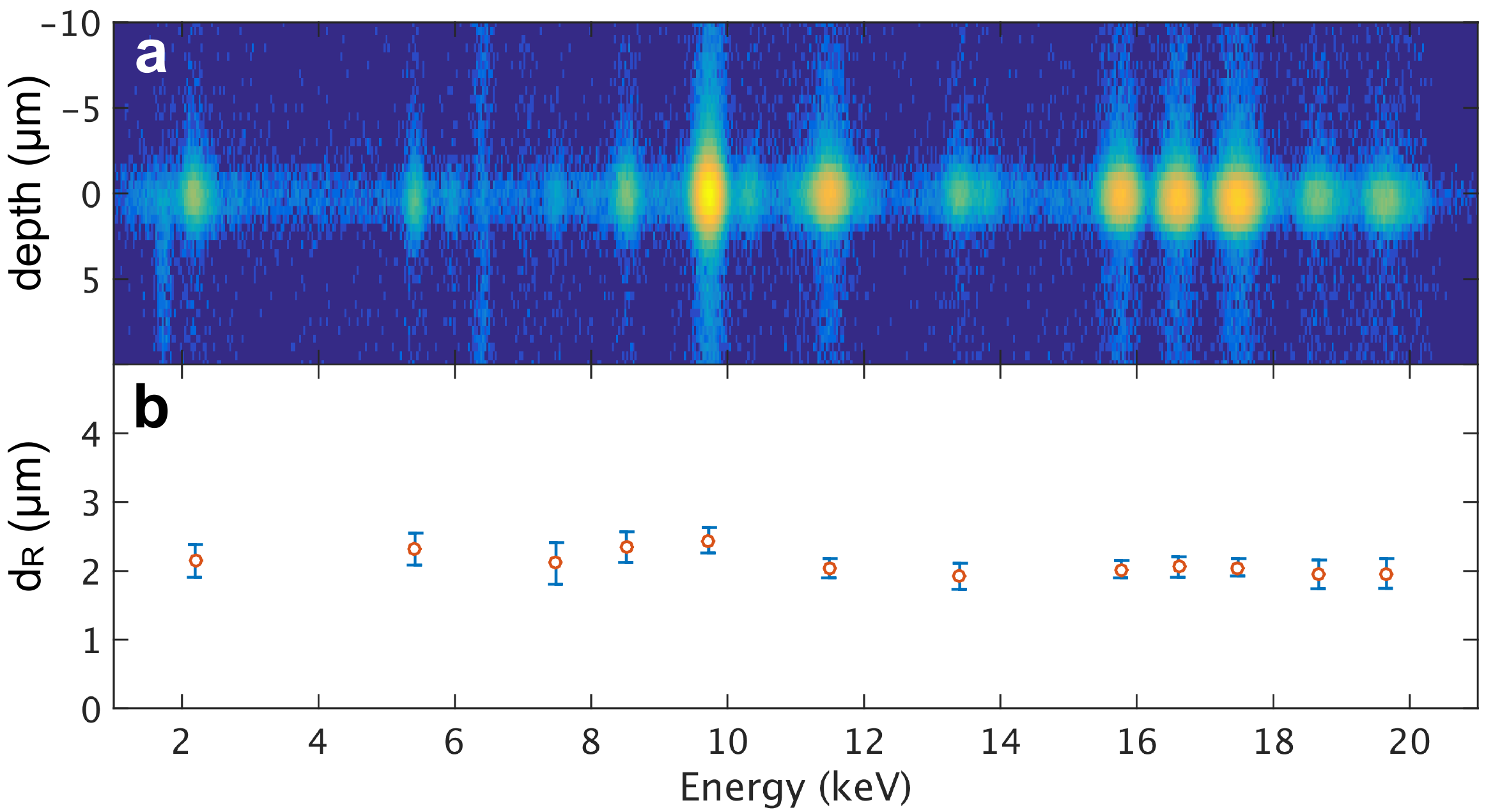}
  \caption{Depth resolution d$_R$, (FWHM) as a function of energy, obtained by fitting different XRF emission line intensities as a function of sample position, as in Figure~\ref{fgr:depth_scans}. Error bars correspond to 95\% confidence intervals. From left to right, the points are obtained from emission lines Au M$\alpha$, Cr K$\alpha$, Ni K$\alpha$ (evidently, Ni is present as a contaminant in the film), Au L1, Au L$\alpha$, Au L$\beta$, Au L$\gamma$, Zr K$\alpha$, Nb K$\alpha$, Mo K$\alpha$, Nb K$\alpha$, and Mo K$\beta$. \label{fig:res_plot}}
\end{figure}

The same scan used to generate Figure~\ref{fgr:depth_scans}b may also be analyzed to demonstrate, as in Refs.~\cite{woll_3d_2012,woll2014}, that the CXRF probe volume using CCA optics is nearly energy independent. Figure~\ref{fig:res_plot}a shows the full spectrum data from the scan shown in Figure~\ref{fgr:depth_scans}b. In the image, each horizontal row of data corresponds to a spectrum obtained with the sample at a particular position in the scan. Vertical features in the image correspond to different emission lines or line groups, most of which arise from the multilayer film. Two exceptions are silicon fluorescence from the substrate at 1.7 keV, and a nearly constant feature at 6.4 keV, corresponding to Fe K$\alpha$ emission and arising from the steel optic holder. Notably absent is fluorescence from germanium. We do, however, expect some Ge background to be present for spectra exhibiting sufficiently high intensity at sufficiently high energy. 

Figure \ref{fig:res_plot}b shows the result of fitting the intensity distributions in Figure~\ref{fig:res_plot}a to a Gaussian function and extracting the best-fit FWHM, which we define as the depth resolution $d_R$. All of the values of $d_R$ are close to 2 \um, and average to 2.1 \umnos. We emphasize that this achromatic behavior in CXRF resolution strongly contrasts that obtained from CXRF setups making use of polycapillaries as the collection optic. We note too that $d_R$ is related to the incident horizontal beam size $a$ and the spatial resolution of the CCA $r$ as \cite{malzer2005model}:
\begin{equation}  \label{eq:dres}
d_R = \sqrt{a^2\sin^2\psi + r^2\cos^2\psi},
\end{equation}
where $\psi$ is the angle between the surface normal and incident beam, 55$^{\circ}$ in the setup employed here. Since $r$ cannot be smaller than the channel width, 2~\umnos, solving this equation for $a$ implies an incident beam size no greater than 2.15~\umnos.

\section{CONCLUSIONS}
This paper summarizes recent success in developing CCA optics as a means of performing direct, 3D XRF microscopy on the micron scale, which has a wide variety of applications spanning the fields of cultural heritage, biology, materials science, geology and many others. Specifically, we have shown that fabricating CCA optics from germanium, rather than silicon, has allowed operation up to 20 keV as opposed to 10 keV, and have demonstrated a spatial resolution of 2 \um from 2-20 keV. We have also provided many details related to the design, fabrication and application of CCA optics, for example relating to the etch procedures, background reduction, and CXRF alignment. CCA optics A and D described here are available in both silicon and germanium for general use at beamline 20ID-B, APS.

\section{ACKNOWLEDGMENTS}
This work was performed in part at the Cornell NanoScale Facility, a member of the National Nanotechnology Coordinated Infrastructure (NNCI), which is supported by the National Science Foundation (Grant ECCS-15420819). This work is also based upon research conducted at the Cornell High Energy Synchrotron Source (CHESS) which is supported by the National Science Foundation and the National Institutes of Health/National Institute of General Medical Sciences under NSF award DMR-1332208. Sector 20 facilities at the Advanced Photon Source, and research at these facilities, are supported by the US Department of Energy - Basic Energy Sciences, the Canadian Light Source and its funding partners, the University of Washington, and the Advanced Photon Source. Use of the Advanced Photon Source, an Office of Science User Facility operated for the U.S. Department of Energy (DOE) Office of Science by Argonne National Laboratory, was supported by the U.S. DOE under Contract No. DE-AC02-06CH11357. S.C. is a Fellow in the Canadian Institutes of Health Research (CIHR) Training grant in Health Research Using Synchrotron Techniques (CIHR- THRUST).\\


\bibliography{ixcom_dna_v2.bbl}%

\end{document}